\documentclass[12pt]{article}

\newcommand{\be}{\begin{equation}}
\newcommand{\ee}{\end{equation}}
\begin{document}
Postal address:
 
{\bf L. Herrera}
 
Apartado 80793, Caracas 1080A, Venezuela, 
 
e-mail: laherrera@telcel.net.ve
 
{\bf N. O. Santos}
 
Centro Brasileiro de Pesquisas F\'{\i}sicas, LAFEX,
 
rua Dr. Xavier Sigaud 150, Urca, CEP 22290-180, 
 
Rio de Janeiro, RJ, Brazil, 
 
e-mail: nos@cbpf.br
 
\title{SHEAR--FREE AND HOMOLOGY CONDITIONS FOR SELF--GRAVITATING
DISSIPATIVE FLUIDS}
\author{L. Herrera$^1$\thanks{Postal
address: Apartado 80793, Caracas 1080A, Venezuela.} \thanks{e-mail:
laherrera@telcel.net.ve} and
N. O. Santos$^{2,3,4}$\thanks{e-mail: nos@cbpf.br; santos@ccr.jussieu.fr}
 \\
{\small $^1$Escuela de F\'{\i}sica, Facultad de Ciencias,}\\
{\small Universidad Central de Venezuela, Caracas, Venezuela.} \\
{\small $^2$LRM--CNRS/UMR 8540, Universit\'e Pierre et Marie Curie, ERGA,}\\
{\small  Bo\^\i te 142, 4 place Jussieu, 75005 Paris Cedex 05, France.}\\
{\small $^3$Laborat\'orio Nacional de Computa\c{c}\~ao Cient\'{\i}fica,}\\
{\small 25651-070 Petr\'opolis--RJ, Brazil.}\\
{\small $^4$ Centro Brasileiro de Pesquisas F\'{\i}sicas, 22290-180
Rio de Janeiro-RJ, Brazil.}\\
}
\maketitle
\begin{abstract}
The shear free condition is studied for dissipative relativistic
self--gravitating fluids in the quasi--static approximation. It is shown
that, in the Newtonian limit, such condition
implies the linear homology law for the velocity of a fluid element,only if
homology conditions are further imposed on the temperature and the
emission rate.It is also shown that the
shear--free plus the homogeneous expansion rate conditions are equivalent
(in the Newtonian limit) to the homology conditions. Deviations from
homology and their prospective
applications to some astrophysical scenarios are discussed, and a model is
worked out.

\vspace{0.3cm} 

{\bf Key words:} gravitation, relativity, hydrodynamics, stellar dynamics,
radiation mechanisms: general, diffusion.
\end{abstract}

\newpage
\section{Introduction}
As it is well known the shear plays an important role in general
relativistic and cosmological models (see Collins \& Wainwright 1983 and Glass 1979
and references therein).
 
In the case of slowly evolving (quasi--static) non--dissipative systems, it
can be shown that in the Newtonian limit, the shear--free condition leads
to the homologous contraction (or
expansion) law for the velocity (Herrera \& Santos 1995). However this is not necessarily the case in the presence of a
heat flow vector (Herrera \& Di Prisco 1996) and/or free streaming radiation (see below).
This fact, and the great relevance of
homology conditions in astrophysics (Schwarzschild 1958; Kippenhahn \& Weigert 1990;
Hansen \& Kawaler 1994)  provide the main motivation for this work.
 
It is our purpose here to explore deeper the link between these two
conditions and  present some astrophysical scenarios where departures from
homologous evolution (keeping the shear--free condition) migth drastically change the whole picture of the system.
 
 Accordingly, we shall consider dissipative systems. Indeed, dissipation
due to the emission of massless
particles (photons and/or neutrinos) is a characteristic process in the
evolution of massive stars. In fact, it seems that the only plausible
mechanism to carry away the bulk of the binding energy of the collapsing
star, leading to a neutron star or black hole is neutrino emission
(Kazanas \& Schramm 1979).
 
In the diffusion approximation, it is assumed that the energy flux of
radiation (as that of
thermal conduction) is proportional to the gradient of temperature. This
assumption is in general very sensible, since the mean free path of
particles responsible for the propagation of energy in stellar
interiors is usually very small as compared with the typical
length of the object.
Thus, for a main sequence star as the sun, the mean free path of
photons at the centre, is of the order of $2\, cm$. Also, the
mean free path of trapped neutrinos in compact cores of densities
about $10^{12} \, g.cm.^{-3}$ becomes smaller than the size of the stellar
core (Arnett 1977; Kazanas 1978).
 
Furthermore, the observational data collected from supernovae 1987A
indicates that the regime of radiation transport prevailing during the
emission process, is closer to the diffusion approximation than to the
streaming out limit (Lattimer 1988).
 
However in many other circumstances, the mean free path of particles
transporting energy may be large enough as to justify the  free streaming
approximation. Therefore we
will include simultaneously both limiting  cases of radiative transport
(diffusion and streaming out), allowing for describing a wide range
situations.
 
As mentioned before homologous evolution, an assumption  widely used in
astrophysics (Schwarzschild 1958; Kippenhahn \& Weigert 1990; 
Hansen \& Kawaler 1994), is known to be equivalent, in the
non--dissipative case, to the shear--free condition in the Newtonian limit
(Herrera \& Santos 1995). As we
shall see here, the presence of dissipative terms requires further the
assumption of homology conditions on temperature and emission rate, in
order to keep  the homologous linear
dependence of the velocity. It will also be shown that imposing the rate of
expansion to be independent of the radial coordinate (together with shear--free condition)
amounts to the full set of homology conditions.
 
Although deviations from the homologous evolution are shown to introduce
extremely small modifications in the expression for the velocity, these
terms might be relevant in some very
specific situations which we will discuss later.
 
It is also worth
mentioning that although the most common method of solving Einstein's
equations is to use comoving coordinates (e.g. May \& White 1966; Wilson 1971;
Burrows \& Lattimer 1986; Adams, Cary \& Cohen 1989; Bonnor \& Knutsen 1993) we shall use noncomoving coordinates, which
implies that
the velocity of any fluid element (defined with respect to a conveniently
chosen set of observers) has to be
considered as a relevant physical variable (Bonnor \& Knutsen 1993).

The plan of the paper is as follows. In Section 2 we define the conventions
and give the field equations and expressions for the kinematical and
physical variables we shall use, in
noncomoving coordinates. In Section 3 we give the general expression for
the velocity and evaluate the dissipative terms. A very simple model is
presented in Section 4. Finally a
discussion of results is presented in Section 5.
 
\section{Relevant  Equations and Conventions}
\subsection{The field equations}
We consider spherically symmetric distributions of collapsing
fluid, which for sake of completeness we assume to be locally anisotropic,
undergoing dissipation in the form of heat flow and/or free streaming
radiation, bounded by a
spherical surface $\Sigma$.
 
The line element is given in Schwarzschild--like coordinates by
 
\begin{equation}
ds^2=e^{\nu} dt^2 - e^{\lambda} dr^2 -
r^2 \left( d\theta^2 + \sin^2\theta d\phi^2 \right),
\label{metric}
\end{equation}
where $\nu(t,r)$ and $\lambda(t,r)$ are functions of their arguments. We
number the coordinates: $x^0=t; \, x^1=r; \, x^2=\theta; \, x^3=\phi$.
The metric (\ref{metric}) has to satisfy Einstein field equations
\begin{equation}
G^\nu_\mu=-8\pi T^\nu_\mu.
\label{Efeq}
\end{equation}
In order to give physical significance to the $T^{\mu}_{\nu}$ components
we apply the Bondi approach (Bondi 1964).
Thus, following Bondi, let us introduce purely locally Minkowski
coordinates ($\tau, x, y, z$)
\begin{eqnarray}
d\tau=e^{\nu/2}dt\,;\qquad\,dx=e^{\lambda/2}dr\,;\qquad\,
dy=rd\theta\,;\qquad\, dz=r\sin\theta d\phi. \nonumber
\end{eqnarray}
Then, denoting the Minkowski components of the energy tensor by a bar,
we have
\begin{eqnarray}
\bar T^0_0=T^0_0\,;\qquad\,
\bar T^1_1=T^1_1\,;\qquad\,\bar T^2_2=T^2_2\,;\qquad\,
\bar T^3_3=T^3_3\,;\qquad\,\bar T_{01}=e^{-(\nu+\lambda)/2}T_{01}. \nonumber
\end{eqnarray}
Next, we suppose that when viewed by an observer moving relative to these
coordinates with proper velocity $\omega$ in the radial direction, the physical
content  of space consists of an anisotropic fluid of energy density $\rho$,
radial pressure $P_r$, tangential pressure $P_\bot$,  radial heat flux
$\hat q$ and unpolarized radiation of energy density $\hat\epsilon$
traveling in the radial direction. Thus, when viewed by this moving
observer the covariant tensor in
Minkowski coordinates is
 
\[ \left(\begin{array}{cccc}
\rho + \hat\epsilon    &  -\hat q - \hat\epsilon  &   0     &   0    \\
-\hat q - \hat\epsilon &  P_r + \hat\epsilon    &   0     &   0    \\
0       &   0       & P_\bot  &   0    \\
0       &   0       &   0     &   P_\bot
\end{array} \right). \]
 
\noindent
Then a Lorentz transformation readily shows that
 
\begin{equation}
T^0_0=\bar T^0_0= \frac{\rho + P_r \omega^2 }{1 - \omega^2} +
\frac{2 Q \omega e^{\lambda/2}}{(1 - \omega^2)^{1/2}} + \epsilon,
\label{T00}
\end{equation}
 
\begin{equation}
T^1_1=\bar T^1_1=-\frac{ P_r + \rho \omega^2}{1 - \omega^2} -
\frac{2 Q \omega e^{\lambda/2}}{(1 - \omega^2)^{1/2}}-\epsilon,
\label{T11}
\end{equation}
 
\begin{equation}
T^2_2=T^3_3=\bar T^2_2=\bar T^3_3=-P_\bot,
\label{T2233}
\end{equation}
 
\begin{eqnarray}
T_{01}=e^{(\nu + \lambda)/2} \bar T_{01}=
-\frac{(\rho + P_r) \omega e^{(\nu + \lambda)/2}}{1 - \omega^2} \nonumber \\
-\frac{Q e^{\nu/2} e^{\lambda}}{(1 - \omega^2)^{1/2}} (1 + \omega^2)
-\epsilon e^{(\nu + \lambda)/2},
\label{T01}
\end{eqnarray}
\noindent
with
 
\begin{equation}
Q \equiv \frac{\hat q e^{-\lambda/2}}{(1 - \omega^2)^{1/2}}
\label{defq}
\end{equation}
and
\begin{equation}
\epsilon\equiv\hat\epsilon\frac{1+\omega}{1-\omega}.
\label{defepsilon}
\end{equation}
 
\noindent
Note that the coordinate velocity in the ($t,r,\theta,\phi$) system, $dr/dt$,
is related to $\omega$ by
 
\begin{equation}
\omega=\frac{dr}{dt}\,e^{(\lambda-\nu)/2}.
\label{omega}
\end{equation}
 
\noindent
Feeding back (\ref{T00}--\ref{T01}) into (\ref{Efeq}), we get
the field equations in  the form
 
\begin{equation}
\frac{\rho + P_r \omega^2 }{1 - \omega^2} +
\frac{2 Q \omega e^{\lambda/2}}{(1 - \omega^2)^{1/2}} +
\epsilon=-\frac{1}{8 \pi}\left[-\frac{1}{r^2}+e^{-\lambda}
\left(\frac{1}{r^2}-\frac{\lambda'}{r} \right)\right],
\label{fieq00}
\end{equation}
 
\begin{equation}
\frac{ P_r + \rho \omega^2}{1 - \omega^2} +
\frac{2 Q \omega e^{\lambda/2}}{(1 -
\omega^2)^{1/2}}+\epsilon=-\frac{1}{8 \pi}\left[\frac{1}{r^2} - e^{-\lambda}
\left(\frac{1}{r^2}+\frac{\nu'}{r}\right)\right],
\label{fieq11}
\end{equation}
 
\begin{eqnarray}
P_\bot = -\frac{1}{8 \pi}\Biggl\{\frac{e^{-\nu}}{4}\left[2\ddot\lambda+
\dot\lambda(\dot\lambda-\dot\nu)\right] \nonumber \\
 - \frac{e^{-\lambda}}{4}
\left(2\nu''+\nu'^2 -
\lambda'\nu' + 2\frac{\nu' - \lambda'}{r}\right)\Biggr\},
\label{fieq2233}
\end{eqnarray}
 
\begin{equation}
\frac{(\rho + P_r) \omega e^{(\nu + \lambda)/2}}{1 - \omega^2} +
\frac{Q e^{\nu/2} e^{\lambda}}{(1 - \omega^2)^{1/2}} (1 + \omega^2)
+\epsilon e^{(\nu + \lambda)/2}=-\frac{\dot\lambda}{8 \pi r},
\label{fieq01}
\end{equation}
where the dots and primes stand for partial derivatives with respect to $t$ nd $r$
respectively.
 
\noindent
Outside of the fluid distribution, the spacetime is that of Vaidya,
given by
 
\begin{equation}
ds^2= \left[1-\frac{2M(u)}{R}\right] du^2 + 2dudR -
 R^2 \left(d\theta^2 + \sin^2\theta d\phi^2 \right),
\label{Vaidya}
\end{equation}
 
\noindent
where $u$ is a coordinate related to the retarded time, such that
$u=constant$ is (asymptotically) a
null cone open to the future and $R$ is a null coordinate ($g_{
R R}=0$). It should
be remarked, however, that strictly speaking, the radiation can be considered
in radial free streaming only at radial infinity.
The two coordinate systems ($t,r,\theta,\phi$) and ($u,
R,\theta,\phi$) are
related at the boundary surface and outside it by
 
\begin{equation}
u=t-r-2M\ln \left(\frac{r}{2M}-1\right),
\label{u}
\end{equation}
 
\begin{equation}
 R=r.
\label{radial}
\end{equation}
 
\noindent
In order to match smoothly the two metrics above on the boundary surface
$r=r_\Sigma(t)$, we first require the continuity of the first fundamental
form across that surface.
Which in our notation implies (Herrera et al. 2002)
\begin{equation}
e^{\nu_\Sigma}=1-\frac{2M}{R_\Sigma},
\label{enusigma}
\end{equation}
\begin{equation}
e^{-\lambda_\Sigma}=1-\frac{2M}{R_\Sigma}.
\label{elambdasigma}
\end{equation}
Where, from now on, subscript $\Sigma$ indicates that the quantity is
evaluated at the boundary surface $\Sigma$ and  $R=R_\Sigma(u)$ is the
equation of the boundary surface in
($u,R,\theta,\phi$)
coordinates.
And
\begin{equation}
\left[P_r\right]_\Sigma=\left[Q\,e^{\lambda/2}\left(1-\omega^2\right)^
{1/2}\right]_\Sigma,
\label{PQ}
\end{equation}
expressing the discontinuity of the radial pressure in the presence
of heat flow, which is a well known result (Santos 1985).
 
Next, it will be useful to calculate the radial component of the
conservation law
\begin{equation}
T^\mu_{\nu;\mu}=0.
\label{dTmn}
\end{equation}
After tedious but simple calculations we get
\begin{equation}
\left(-8\pi T^1_1\right)'=\frac{16\pi}{r} \left(T^1_1-T^2_2\right)
+ 4\pi \nu' \left(T^1_1-T^0_0\right) +
\frac{e^{-\nu}}{r} \left(\ddot\lambda + \frac{\dot\lambda^2}{2}
- \frac{\dot\lambda \dot\nu}{2}\right),
\label{T1p}
\end{equation}
which in the static case becomes
\begin{equation}
P'_r=-\frac{\nu'}{2}\left(\rho+P_r\right)+
\frac{2\left(P_\bot-P_r\right)}{r},
\label{Prp}
\end{equation}
representing the generalization of the Tolman--Oppenheimer--Volkof equation
for anisotropic fluids (Bowers \& Liang 1974).
 
\subsection{The kinematical variables}
The components of the shear tensor are defined by
\begin{equation}
\sigma_{\mu\nu}=u_{\mu;\nu}+u_{\nu;\mu}-u_{\mu}a_{\nu}-u_{\nu}a_{\mu}-
\frac{2}{3}{\Theta}P_{\mu\nu},
\label{shear}
\end{equation}
where
\begin{equation}
P_{\mu\nu}=g_{\mu\nu}-u{_\mu}u_{\nu}\,;\qquad\,
\Theta=u^{\mu}_{;\mu}\,;\qquad\,a_{\mu}=u^{\nu}u_{\mu;\nu},
\label{P,Theta,a}
\end{equation}
denote, $P_{\mu\nu}$ the projector onto the three space orthogonal to $u^\mu$, $\Theta$  the
expansion and $a_{\mu}$ the acceleration. 
 
A simple calculation gives
\begin{eqnarray}
\Theta=\frac{e^{-\nu/2}}{2\left(1-\omega^2\right)^{1/2}}
\left(\dot\lambda + \frac{2\omega\dot\omega}{1-\omega^2}\right) \nonumber \\
+\frac{e^{-\lambda/2}}{2\left(1-\omega^2\right)^{1/2}}
\left(\omega\nu' + 2\omega' +
\frac{2\omega^2\omega'}{1-\omega^2} +
\frac{4\omega}{r}\right),
\label{Theta}
\end{eqnarray}
 
\begin{eqnarray}
\sigma_{11}= -\frac{2}{3\left(1-\omega^2\right)^{3/2}}
\left[e^{\lambda}e^{-\nu/2}\left(\dot\lambda +
\frac{2\omega\dot\omega}{1-\omega^2}\right) \right. \nonumber \\
\left. +e^{\lambda/2}
\left(\omega\nu' + \frac{2\omega'}{1-\omega^2}
- \frac{2\omega}{r}\right)  \right],
\label{shear11}
\end{eqnarray}
 
\begin{equation}
\sigma_{22}=-\frac{e^{-\lambda} r^2 \left(1-\omega^2\right)}{2}
\sigma_{11},
\label{shear22}
\end{equation}
 
\begin{equation}
\sigma_{33}=-\frac{e^{-\lambda} r^2 \left(1-\omega^2\right)}{2}
\sin^2{\theta} \sigma_{11},
\label{shear33}
\end{equation}
 
\begin{equation}
\sigma_{00}=\omega^2 e^{-\lambda} e^\nu \sigma_{11},
\label{shear00}
\end{equation}
 
\begin{equation}
\sigma_{01}=-\omega e^{\left(\nu-\lambda\right)/2} \sigma_{11},
\label{shear01}
\end{equation}
and for the shear scalar, $\sigma=(\sigma_{\mu\nu}\sigma^{\mu\nu})^{1/2}$,
\begin{equation}
\sigma =\sqrt{3}\left[ \frac \Theta 3-\frac{e^{-\lambda /2}}r\frac
{\omega}{(1-\omega ^2)^{1/2}}\right]. \label{eq:esf}
\end{equation}
 
\subsection{The Weyl tensor}
The model to be presented in Section 4 is obtained from the assumption of
conformal flatness. Furthermore, since the publication of Penrose`s (1979) work,
there has been an
increasing interest in studying the possible role of Weyl tensor (or some
function of it)
in the evolution of self-gravitating systems (Wainwright 1984; Goode \& Wainwright 1985;
Bonnor 1985, 1987; Goode, Coley \& Wainwright 1992; Pelavas \& Lake 2000 ; 
Herrera et al. 2001). This interest is
reinforced by the fact that for spherically symmetric distribution of
fluid, the Weyl tensor may be defined
exclusively in terms of the density contrast and the local anisotropy  of
the pressure (see below), which in turn are known to affect the fate of
gravitational collapse (Mena \& Tavakol 1999; Eardley \& Smarr 1979; Christodoulou 1984;
Newman 1986; Waugh \& Lake 1988; Dwivedi \& Joshi 1992; Singh \& Joshi 1996; 
Herrera \& Santos 1997; Bondi 1993; Barreto 1993; Coley \& Tupper 1994; 
Martinez, Pavon \& Nunez 1994; Singh, Singh \& Helmi 1995; Das, Tariq \& Biech 1995;
Maartens, Maharaj \& Tupper 1995; Das et al. 1997; Corchero 1998a,b; Bondi 1999;
Hernandez, Nunez \& Percoco 1999; Harko \& Mak 2000; Das \& Kloster 2000; 
Joshi, Dadhich \& Maartens 2002; Krisch \& Glass 2002; Corchero 2002; Harko \& Mak 2002;
Mak \& Harko 2002). 
 
Therefore it is worthwhile to include here some expressions for the Weyl tensor.
Thus, using Maple V, it is found
that all non--vanishing components of the Weyl tensor are
proportional to
 
\begin{equation}
W \equiv \frac{r}{2} C^{3}_{232}=W_{(s)} + \frac{r^3 e^{-\nu}}{12}
\left(\ddot\lambda + \frac{\dot\lambda^2}{2} -
\frac{\dot\lambda \dot\nu}{2}\right),
\label{W}
\end{equation}
where
\begin{equation}
W_{(s)} =
\frac{r^3 e^{-\lambda}}{6}
\left( \frac{e^\lambda}{r^2} - \frac{1}{r^2} +
\frac{\nu' \lambda'}{4} - \frac{\nu'^2}{4} -
\frac{\nu''}{2} - \frac{\lambda'}{2r} + \frac{\nu'}{2r} \right),
\label{Ws}
\end{equation}
corresponds to the contribution in the static (and quasi--static) case.
Also, from the field equations and the definition of the Weyl tensor it can
be easily shown that (see Herrera et al. 1998 for details)
\begin{equation}
W = - \frac{4 \pi}{3} \int^r_0{r^3 \left(T^0_0\right)' dr} +
\frac{4 \pi}{3} r^3 \left(T^2_2 - T^1_1\right).
\label{Wint}
\end{equation}
 
\subsection{The slowly evolving approximation}
In this work we shall consider exclusively slowly evolving systems. That
means that our sphere changes slowly on a time scale that is very long
compared to the typical time in which it reacts on a slight perturbation of
hydrostatic equilibrium, this typical time is called hydrostatic time scale.
Thus our system is always in hydrostatic equilibrium (very close to) and
its evolution may be regarded as a sequence of static models linked by
(\ref{fieq01}).
 
This assumption is very sensible because the hydrostatic time scale is
very small for almost any phase of the life of a star. It is of the
order of 27 minutes for the sun, 4.5 seconds for a white dwarf and
$10^{-4}$ seconds for a neutron star of one solar mass and $10Km$
radius (Schwarzschild 1958; Kippenhahn \& Weigert 1990; Hansen \& Kawaler 1994).
 
Let us now express this assumption through conditions for $\omega$ and
metric functions.
 
First of all, slow contraction (or expansion) means that the radial
velocity $\omega$ measured by the Minkowski observer, as well as time
derivatives are so small that their products as well as second time
derivatives can be neglected. Thus we shall assume
\begin{equation}
\ddot\nu\approx\ddot\lambda\approx\dot\lambda \dot\nu\approx
\dot\lambda^2\approx\dot\nu^2\approx
\omega^2\approx\dot\omega\approx 0.
\end{equation}
Then, it follows from (\ref{fieq01}) that $Q$ and
$\epsilon$ are, at most, of order $O(\omega)$. Henceforth, with this
approximation, (\ref{T1p}) becomes
\begin{equation}
(P_r+\epsilon)' + \left(\rho + P_r + 2\epsilon \right) \frac{\nu'}{2} - 2
\frac{P_\bot-P_r-\epsilon}{r} = 0
\label{equi}
\end{equation}
which is the equation of hydrostatic equilibrium for an anisotropic fluid
radiating a null fluid of energy density $\epsilon$.
 
Thus, as mentioned before, the system, although evolving, is in
hydrostatic equilibrium (up to order $O(\omega)$), this allows for a
very simple extension of any static solution to the slowly evolving case.
 
\section{Shear--free and homology conditions}
As mentioned before the only relevant component of the shear tensor is
$\sigma_{11}$ given by equation (\ref{shear11}). Evaluating this last equation in the
slowly evolving  approximation, we obtain
\begin{equation}
\sigma_{11}=-\frac{2}{3} e^{\lambda}\left[e^{-\nu/2}\dot \lambda
+e^{-\lambda/2}\left(\omega \nu'+2\omega'-\frac{2\omega}{r}\right)\right].
\label{shear11prev}
\end{equation}
Next, using (\ref{fieq01}) and
\begin{equation}
P_r+\rho=\frac{e^{-\lambda}}{8\pi r}(\nu'+\lambda')-2\epsilon,
\label{inter}
\end{equation}
easily obtained from (\ref{fieq00}) and (\ref{fieq11}), one gets
\begin{eqnarray}
\sigma_{11}=-\frac{2\sigma_{22}}{r^2}e^{\lambda}=
-\frac{2\sigma_{33}}{r^2 \sin^2\theta}e^{\lambda} \nonumber \\
=-\frac{4}{3} e^{\lambda/2}\left(\omega' - \frac{\omega\lambda'}{2}
-\frac{\omega}{r}-4{\pi}rQ{e^{3\lambda/2}-4{\pi}r\epsilon e^{\lambda}}\right).
\label{shear11f}
\end{eqnarray}
We can solve (\ref{shear11f}) for $\omega$, to obtain
\begin{equation}
\omega=\omega_\Sigma \frac{r}{r_\Sigma}
e^{\left(\lambda-\lambda_\Sigma\right)/2}
- 4\pi r e^{\lambda/2} \int_r^{r_\Sigma}{\left(Q e^\lambda +\epsilon
e^{\lambda/2} -\frac{3}{16 \pi}
e^{-\lambda} \frac{\sigma_{11}}{r}\right)dr}.
\label{intsig11}
\end{equation}
From the above equation we find that in the non--dissipative, shear--free
case we obtain
\begin{equation}
\omega=\omega_{\Sigma}\frac{r}{r_\Sigma}{e^{(\lambda-\lambda_\Sigma)/2}}.
\label{velocidad}
\end{equation}
While in the Newtonian limit we have $M(u)\approx\lambda\approx\nu\approx0$
and  we recover the well known linear expression, typical  of the
homologous evolution (Schwarzschild 1958; Kippenhahn \& Weigert 1990; Hansen \& Kawaler 1994),
\begin{equation}
\omega_{Newt}=\omega_{\Sigma}\frac{r}{r_{\Sigma}}.
\label{omeganewt}
\end{equation}
Also, from (\ref{eq:esf}) evaluated in the slowly evolving approximation,
it follows that in the shear--free motion
\begin{equation}
\Theta=\frac{3 \omega}{r}e^{-\lambda/2},
\label{expansion}
\end{equation}
which of course is valid also in the dissipative case. Using
(\ref{velocidad}), we can write
\begin{equation}
\Theta=\frac{3 \omega_{\Sigma}}{r_{\Sigma}}e^{-\lambda_{\Sigma}/2},
\label{expansionI}
\end{equation}
implying that even in the general (relativistic) case, the expansion rate
is homogeneous  (independent of $r$) for the slow, and dissipativeless
shear--free motion.
 
Let us now consider the dissipative shear--free case.
\par 
From the relativistic Maxwell-Fourier law, we have
\begin{equation}
q^\mu=\kappa{P^{\mu\nu}}\left(T,_\nu-Ta_\nu\right),
\label{qmuT}
\end{equation}
or
\begin{equation}
q^1=Q=-\kappa e^{-\lambda}\left(T'+\frac{T\nu'}{2}\right),
\label{q1T}
\end{equation}
where $T$ is the temperature and $\kappa$ denotes the coefficient
of conduction. It should be reminded that in the quasi--static
approximation, the system is assumed to be relaxed at all times (the
relaxation time is zero) and
accordingly, any hyperbolic transport equation reduces to (\ref{qmuT}).
 
 Then feeding back (\ref{q1T}) into (\ref{intsig11}) and using
(\ref{elambdasigma}) together with the shear--free condition, we obtain
\begin{eqnarray}
\omega=\left[\omega_{\Sigma}\frac{r}{r_{\Sigma}}
\left(1-\frac{2M(u)}{r_\Sigma}\right)^{1/2}+
4{\pi}{\kappa}\left(T_\Sigma-T\right)r \right. \nonumber \\
\left. +2{\pi}{\kappa}r
\int_r^{r_\Sigma}{T\nu'dr}- 4\pi r \int_r^{r_\Sigma}{\epsilon
e^{\lambda/2}dr}\right]e^{\lambda/2},
\label{otravelo}
\end{eqnarray}
which in  the Newtonian limit yields
\begin{equation}
\omega_{Newt}=\omega_{\Sigma}\frac{r}{r_{\Sigma}}+4\pi\kappa\left(T_{\Sigma}-T\right)r 
-4\pi r\int_r^{r_\Sigma}{\epsilon dr}.
\label{omeganewt}
\end{equation}
Also, it follows  that  the expansion (47) with (51) can be written
\begin{equation}
\Theta=\Theta_{\Sigma}+3 \left[
4{\pi}{\kappa}\left(T_\Sigma-T\right)+2{\pi}{\kappa}
\int_r^{r_\Sigma}{T\nu'dr}- 4\pi \int_r^{r_\Sigma}{\epsilon
e^{\lambda/2}dr}\right].
\label{otratheta}
\end{equation}
 
Thus, unlike the non-dissipative case (see also Herrera \& Di Prisco 1996), the shear-free
collapse in the Newtonian limit does not yield the linear law of homologous
contraction (Schwarzschild 1958; Kippenhahn \& Weigert 1990; Hansen \& Kawaler 1994), 
unless we impose further homology conditions on
$T$ and $\epsilon$, i.e. unless we assume that for any given fluid element,
all along the evolution
$$
\frac{T}{T_{\Sigma}}=\mbox{constant},
$$
$$
\frac{\epsilon}{\epsilon_{\Sigma}}=\mbox{constant}.
$$
From (47) we observe that  the sign of $\omega$ for any value of $r$, is
not necessarily
the same as that of $\omega_\Sigma$ (as is the case in the non-dissipative
evolution).
In particular,  for sufficiently large (negative) gradient of temperature
and/or sufficiently large (positive) $\epsilon$ term, we
may have $\omega_\Sigma>0$ and $\omega<0$. The same conclusion of course
applies to $\Theta$.
 
In other words the system may be evolving in such a way that inner
shells collapse, whereas outer ones expand.
 
This effect, which we have called  ``thermal peeling'' (Herrera \& Di Prisco 1996), 
is also present in the
relativistic regime, provided the third term in the right side of
(\ref{otravelo}) is not too large. It represents the analog of the
``cracking'', however whereas the later takes place, under some conditions,
when the system abandons the state of
equilibrium or quasi--equilibirum (Herrera 1992), the former occurs while
the systems is evolving quasi--statically.
 
However, observe that expressing variables in  c.g.s. units, we have that,
$$
\kappa T \sim 10^{-59} \, [\kappa] [T] cm^{-1},
$$
where $[\kappa]$ and $[T]$ denote the numerical values of these quantities
as measured in 
\linebreak
$erg \, s^{-1} \, cm^{-1} \, K^{-1}$ and $K$ respectively.
Therefore extremely high conductivities and/or $\triangle T$ are required
for thermal peeling to be observed in the Newtonian regime.
Also, we have
 
$$
\epsilon  \sim 10^{-59} \, [\epsilon] cm^{-2},
$$
where $[\epsilon]$  denotes the numerical values of this quantity
as measured in $erg \, s^{-1} \, cm^{-2}$.
 
Before closing this section, it is worth mentioning  that, in general, such
high thermal conductivities are associated to highly compact, degenerate
objects where the Newtonian limit is not reliable.
 
Also, it should be noticed that in (\ref{omeganewt}) it has been
assumed that terms of order $O(M/r_\Sigma)$ and higher are negligible
with respect to $\kappa \left(T_\Sigma - T\right)$. This of course is
not always true, as commented above, in which case (\ref{omeganewt})
is not valid.
Finally, it is worth noticing that demanding $\Theta$ to be homogeneous, we
are lead to the homologous contraction, implying thereby that (in the
Newtonian limit), the shear--free and homogeneous 
expansion rate conditions are equivalent to the whole
set of homologous conditions.
 
\section{A model}
In order  to illustrate the point raised in Section 3, let us present a
very simple model based on the assumption of conformal flatness and
shear--free condition. Also, since local
anisotropy does not enter explicitly in (\ref{otravelo}) we shall assume
$P_{r}=P_{\bot}$.
 
Thus assuming $W=0$, it follows from (\ref{Wint})
\begin{equation}
\rho'=\frac{3\epsilon}{r}.
\label{epsmodel}
\end{equation}
Next, taking for simplicity $Q=0$ (pure free streaming dissipation) and
\begin{equation}
\epsilon=\beta r \left(1-\frac{r}{r_{\Sigma}}\right),
\label{ep}
\end{equation}
with $\beta=\beta(t)$ one obtains
\begin{equation}
\rho=3\beta r \left(1-\frac{r}{2 r_{\Sigma}}\right)+\gamma(t),
\label{rhomodel}
\end{equation}
where $\gamma=\rho(0,t)$.
Observe that with this choice of $\epsilon$ (if we assume $Q=0$, i.e the
dissipation takes place at the free streaming approximation exclusively),
the evolution proceeds adiabatically
(the total mass is constant) even though
$\epsilon
\neq 0$ within the sphere.
 
Next, from the definition of the mass function (Herrera \& Santos 1995)
\begin{equation}
m(r,t) = 4 \pi\int^r_0{r^2 T^0_0 dr} =\frac{r}{2}(1-e^{-\lambda})
\label{mass}
\end{equation}
and junction conditions, it follows
\begin{equation}
m= \frac{4 \pi  r^3 }{3}\left(3\beta r+\gamma -\frac{3 \beta r^2}{2
r_{\Sigma}}\right),
\label{masmodel}
\end{equation}
and
\begin{equation}
\beta= \frac{2}{3 r_{\Sigma}}\left(\frac{3M}{4\pi r_{\Sigma}^3}-\gamma\right),
\label{masmodel}
\end{equation}
with $M=m_{\Sigma}$.
As expected, if we put $\beta=0$ we recover the well known interior
Scwarzshild solution (evolving quasi--statically).
The remaining of the metric and physical variables may now be easily
obtained from the field equations.
Feeding back (\ref{ep}) into (\ref{intsig11}) one sees that playing with
$\beta$ it is possible (at least in principle) to obtain $\omega<0$ for
some values of $r$, even though
$\omega_{\Sigma}$ is assumed possitive (peeling).
 
\section{Conclusions}
We have seen how dissipative terms affects the radial dependence of
$\omega$ and the expansion rate, in the shear--free case, if we relax the
homology conditions on dissipative
variables. We have also seen that the dissipative terms may lead to a
``peeling''. However  these contributions appear to be extremely small and
therefore it is pertinent  to ask if
there exist  astrophysical scenarios where dissipative contributions  might
have some effect on
$\omega$, and in particular if they could produce a ``peeling''.
 
Assuming the highest values for luminosity at the last stages of stellar
evolution, of the order of $10^5$ times the  sun luminosity, produced at a
shell of radius of $1/10$ of
solar radius, we only get
$$
\epsilon  \sim 10^{-36} \, cm^{-2}.
$$
 
A more promising case is provided by the Kelvin--Helmholtz phase of the
birth of a neutron star (Burrows \& Lattimer 1986). Indeed in this phase, during tens of
seconds, some $10^{53}$ ergs are
radiated away. If this energy is transported via diffusion to the surface,
then assuming (Flowers \& Itoh 1979,1981)
\begin{equation}
\kappa \approx 10^{23}[\rho/10^{14} g \, cm^{-3}]
[10^{8} K/T] erg \, s^{-1} \, cm^{-1} \, K^{-1}, \nonumber
\label{kap}
\end{equation}
we see that the corresponding contribution to (\ref{otravelo}) is still too
small. However if we assume that part of $10^{53}$ ergs are propagated  in
the free streaming regime, then the last term in (\ref{otravelo}) for
sufficiently small $r$ (as
compared to $r_{\Sigma}$) is of the order of
$10^{52}/r_{\Sigma}$. Therefore for positive surface velocities
of the order of $30 m/s$  there may be a peeling ($\omega<0$ for
$r<r_{\Sigma}$).
 
Finally, let us mention that in a pre--supernovae event, values of the
order of $10^{13}$ and $10^{37}$ have been estimated for $[T]$ and
$[\kappa]$ respectively (Martinez 1996). With
these values, it is clear that peeling is also possible, in particular for
sufficiently large values of $r_{\Sigma}$.
 
\section*{References}
Adams R., Cary B., Cohen J., 1989, Ap\&SS. 155, 271
 
\noindent
Arnett W. D., 1977, ApJ. 218, 815
 
\noindent
Barreto W., 1993, Ap\&SS. 201, 191
 
\noindent
Bondi H., 1964, Proc. R. Soc. London A 281, 39
 
\noindent
Bondi H., 1993, MNRAS. 262, 1088
 
\noindent
Bondi H., 1999, MNRAS. 302, 337
 
\noindent
Bonnor W. B., 1985, Phys. Lett. A 112, 26
 
\noindent 
Bonnor W. B., 1987, Phys. Lett. A 122, 305
 
\noindent
Bonnor W. B., Knutsen H., 1993, Int. J. Theor. Phys. 32, 1061
 
\noindent
Bowers R., Liang E., 1974, ApJ. 188, 657
 
\noindent
Burrows A., Lattimer J., 1986, ApJ. 307, 178
 
\noindent
Christodoulou D., 1984, Commun. Math. Phys. 93, 171
 
\noindent
Coley A., Tupper B., 1994, Class. Quantum Grav. 11, 2553
 
\noindent
Collins C. B., Wainwright J., 1983, Phys. Rev. D 27, 1209
 
\noindent
Corchero E., 1998a, Class. Quantum Grav. 15, 3645
 
\noindent
Corchero E., 1998b, Ap\&SS. 259, 31
 
\noindent
Corchero E., 2002, Class. Quantum Grav. 19, 417
 
\noindent
Das A., Kloster S., 2000, Phys. Rev. D  62, 104002
 
\noindent
Das A., Tariq N., Aruliah D., Biech T., 1997, J. Math. Phys. 38, 4202
 
\noindent
Das A., Tariq N., Biech T., 1995, J. Math. Phys. 36, 340
 
\noindent
Dwivedi I. H., Joshi P. S., 1992, Class. Quantum Grav. 9, L69
 
\noindent
Eardley D. M., Smarr L., 1979, Phys. Rev. D 19, 2239
 
\noindent
Flowers N., Itoh N., 1979, ApJ. 230, 847
 
\noindent
Flowers N., Itoh N., 1981, ApJ. 250, 750
 
\noindent
Goode S. W., Coley A., Wainwright J., 1992, Class. Quantum Grav. 9, 445
 
\noindent
Glass E. N., 1979, J. Math. Phys. 20, 1508
 
\noindent
Goode S. W., Wainwright J., 1985, Class. Quantum Grav. 2, 99
 
\noindent
Hansen C., Kawaler S., 1994, Stellar Interiors: Physical Principles, Structure and
Evolution, Springer Verlag, Berlin
 
\noindent
Harko T., Mak M. K., 2000, J. Math. Phys. 41, 4752
 
\noindent 
Harko T., Mak M. K., 2002, Ann. Phys. (Leipzig) 11, 3
 
\noindent
Hernandez H., Nunez L., Percoco U., 1999, Class. Quantum Grav. 16, 897
 
\noindent
Herrera L., 1992, Phys. Lett A 165, 206
 
\noindent
Herrera L., Barreto W., Di Prisco A., Santos N. O., 2002, Phys. Rev. D 65, 104004
 
\noindent
Herrera L., Di Prisco A., 1996, Phys. Rev. D 55, 2044
 
\noindent
Herrera L., Di Prisco A., Hernandez-Pastora J., Santos N. O., 1998, Phys. Lett. A 237, 113 
 
\noindent
Herrera L., Di Prisco A., Ospino J., Fuenmayor E., 2001, J. Math. Phys. 42, 2129
 
\noindent
Herrera L., Santos N. O., 1995, Gen. Rel. Grav. 27, 1071
 
\noindent
Herrera L., Santos N. O., 1997, Phys. Rep. 286, 53
 
\noindent
Joshi P. S., Dadhich N., Maartens R., 2002, Phys. Rev. D 65, 101501
 
\noindent
Joshi P. S., Dwivedi I. H., 1993, Phys. Rev. D 47, 5357
 
\noindent
Kazanas D., 1978, ApJ. 222, 2109
 
\noindent
Kazanas D., Schramm D., 1979, in Smarr L., ed., Sources of Gravitational Radiation,
Cambridge University Press, Cambridge
 
\noindent
Kippenhahn R., Weigert A., 1990, Stellar Structure and Evolution, Springer Verlag, Berlin
 
\noindent
Krisch J., Glass E., 2002, J. Math. Phys. 43, 1509
 
\noindent
Lattimer J., 1988, Nucl. Phys. A 478, 199
 
\noindent
Maartens R., Maharaj S., Tupper B., 1995, Class. Quantum Grav. 12, 257
 
\noindent
Mak M. K., Harko T., 2002, Chinese J. Astron. Astrophys. 2, 248
 
\noindent
Martinez J., 1996, Phys. Rev. D 53, 6921
 
\noindent
Martinez J., Pavon D., Nunez L., 1994, MNRAS. 271, 463
 
\noindent
May M., White R., 1966, Phys. Rev. D 141, 1232
 
\noindent
Mena F., Tavakol R., 1999, Class. Quantum Grav. 16, 435
 
\noindent
Newman R. P. A. C., 1986, Class. Quantum Grav. 3, 527
 
\noindent
Pelavas N., Lake K., 2000, Phys. Rev. D 62, 044009
 
\noindent
Penrose R., 1979, in Hawking S. W., Israel W., ed., An Einstein Centenary Survey, 
Cambridge University Press, Cambridge
 
\noindent
Santos N. O., 1985, MNRAS. 216, 403
 
\noindent
Schwarzschild M., 1958, Structure and Evolution of the Stars, Dover, New York
 
\noindent
Singh T. P., Joshi P. S., 1996, Class. Quantum Grav. 13, 559
 
\noindent
Singh T., Singh G. P., Helmi A. M., 1995, Il Nuovo Cimento B 110, 387
 
\noindent
Wainwright J., 1984, Gen. Rel. Grav. 16, 657
 
\noindent
Waugh B., Lake K., 1988, Phys. Rev. D 38, 1315
 
\noindent
Wilson J., 1971, ApJ. 163, 209

\end{document}